\documentclass[12pt]{iopart}
\usepackage{graphicx}
\usepackage{hyperref}
\usepackage{cite}
\begin{document}

\title[Trajectory-based models in strong-field physics]{Trajectory-based models in strong-field physics}

\author{N.~I.~Shvetsov-Shilovski$^{1,2}$}

\address{$^1$Rudolf Bembenneck Gesamtschule Burgdorf, Vor dem Celler Tor 50, 31303 Burgdorf, Germany}
\address{$^2$Institut f\"{u}r Theoretische Physik, Leibniz Universit\"{a}t Hannover, 30167 Hannover, Germany }
\ead{n79@narod.ru}
\vspace{10pt}
\begin{indented}
\item[]October 2025
\end{indented}

\begin{abstract}
We review various semiclassical models for strong-field physics. These semiclassical models employ ensembles of classical trajectories to simulate electron motion in the continuum after being released from an atom or molecule by an external laser field. We discuss different approaches to trajectory-based simulations and identify their advantages and shortcomings. We also review some of the recent applications of semiclassical models to the key strong-field phenomena: above-threshold ionization, high-order harmonic generation, nonsequential double ionization, and frustrated tunneling ionization. 
\end{abstract}

\vspace{2pc}
\noindent{\it Keywords}: strong-field physics, semiclassical models, above-threshold ionization, generation of high-order harmonics, nonsequential double ionization, frustrated tunneling ionization
%

\submitto{Laser Phys.}
%

%
 
%

\section{Introduction}\label{intro}

Strong-field physics explores highly nonlinear processes that arise when atoms and molecules are exposed to intense laser fields. Among the most extensively studied effects are above-threshold ionization (ATI), the emergence of a high-energy plateau in the photoelectron spectrum (high-order ATI), high-order harmonic generation (HHG), and nonsequential double ionization (NSDI) [see Refs.~\cite{FedorovBook1997,DeloneBook2000,BeckerRev2002,MilosevicRev2003,FaisalRev2005,FariaRev2011,BeckerRev2012,KitzlerBook2016,LinBook2018} for reviews].  These complex phenomena require a robust theoretical description, and the theoretical approaches used in strong-field physics are continually being refined. These approaches are based on the direct numerical solution of the time-dependent Schr\"{o}dinger equation (see, e.g., Refs.~\cite{Muller1999,Bauer2006,Madsen2007,Patchkovski2016}), the strong-field approximation (SFA) \cite{Keldysh1964,Faisal1973,Reiss1980}, and semiclassical models employing classical trajectories to describe electron motion in the continuum. Widely known examples of semiclassical models are the two-step model \cite{Linden1988,Gallagher1988,Corkum1989} and the three-step model \cite{Krause1992,Corkum1993}.

In many cases, direct numerical solutions of the time-dependent Schr\"{o}dinger equation (TDSE) in the single-active electron approximation (see Refs.~\cite{Kulander1988,Kulander1991}) yield results that are in good agreement with experimental observations. However, the numerical wavefunction alone often provides limited insight into the mechanisms underlying the phenomena of interest. Moreover, the capabilities of modern computer clusters impose practical limitations on these approaches. A particularly illustrative example is the strong-field ionization of molecules. While solving the TDSE in three spatial dimensions is, in principle, possible, it remains computationally feasible only for the simplest molecular systems and typically requires the selection of the most relevant degrees of freedom \cite{Palacios2006,Saenz2014}. Molecular ionization in strong laser fields is substantially more complex than its atomic counterpart due to several factors: the presence of additional nuclear degrees of freedom, the interplay of multiple timescales, and the intricate spatial structure of molecular orbitals. Under typical experimental conditions, nuclear motion cannot be neglected and must be treated on an equal footing with the laser-induced electronic dynamics. Furthermore, the different nuclear configurations of molecules give rise to orbitals with diverse symmetries, further enriching the dynamics and complicating theoretical treatment.

In the strong-field approximation, ionization is modelled as a direct transition from an initial bound state, which is assumed to be unaffected by the laser field, to a Volkov state - the wavefunction of a free electron in an electromagnetic field. This approach neglects both intermediate bound-state dynamics and the influence of the Coulomb potential on the electron motion in the continuum. The SFA provides a clear and simple picture of many processes in strong laser fields and, importantly, often permits analytical solutions. However, the approximations used in the SFA are significant and can lead to qualitatively incorrect predictions. A well-known example is the fourfold symmetry in the photoelectron angular distributions predicted by the SFA for ionization by elliptically polarized laser fields \cite{PPT2}. In contrast, the experimentally measured distributions possess only inversion symmetry, i.e., they are asymmetric in each half of the polarisation plane \cite{Bashkansky1988}. Theoretical investigations \cite{Basile1988,Lambropoulos1988,Muller1988,Kis1991,Jaron1999,Manakov2000,Shvetsov2004} have shown that this discrepancy arises from the neglect of the Coulomb interaction between the electorn and the parent ion in the continuum, which plays an important role in shaping the electron dynamics after ionization.

Presently, semiclassical models are widely used for studies of various processes in strong laser fields. This is due to several advantages inherent to trajectory-based approaches. First of all, these approaches provide valuable physical insight into strong-field phenomena in terms of electron trajectories in an external field. Therefore, we briefly discuss the dynamics of a ionized electron in a strong laser field.  In above-threshold ionization (ATI), an electron absorbs more photons than the minimum number required for ionization. Investigations of the ATI process have shown that the majority of electrons do not undergo hard recollisions with their parent ions after being liberated by a laser pulse. These electrons, commonly referred to as direct electrons, contribute predominantly to the low-energy part of the ATI energy spectrum, i.e., they have energies $E<2U_{p}$, where $U_{p}=F^{2}/4\omega^2$ is the ponderomotive energy (atomic units are used throughout the paper unless indicated otherwise). Here, $F$ and $\omega$ are the amplitude and the frequency of the external field, respectively. The direct electron spectrum can be described with the two-step model \cite{Linden1988,Gallagher1988,Corkum1989}. In the first step of the two-step model, the electron tunnels out of the atom through the potential barrier, and in the second step it propagates along a classical trajectory in the continuum driven by the laser field.

\begin{figure}[h]
\centering
\includegraphics[width=0.5\textwidth]{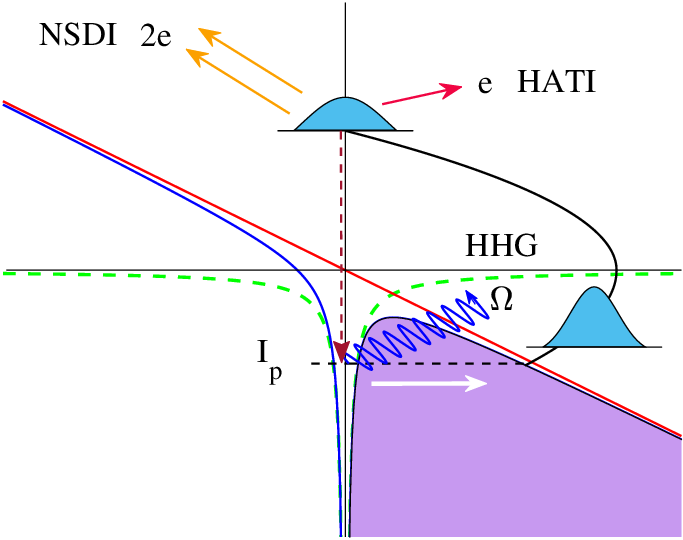}
\caption{The three-step model of strong-field ionization involves (i) electron tunneling through the effective potential barrier, (ii) acceleration by the laser field, and (iii) recollision with the parent ion, which can lead to high-order above-threshold ionization (via elastic scattering), high-harmonic generation (via recombination), or nonsequential double ionization (via the liberation of a second electron). The effective potential barrier (solid blue curve) results from the superposition of the Coulomb potential (dashed green curve) and the laser field potential (solid red curve).}
\label{three_step}
\end{figure}

Along with direct electrons, there are also electrons that are driven back to their parent ions by the oscillating laser field. These rescattered electrons can undergo large-angle scattering (up to $180^{\circ}$) on the parent ions. The rescattered electrons are responsible for the high-energy plateau observed in the ATI spectrum. The concept of rescattering has also allowed to understand the physical mechanism of the HHG and NSDI processes. In the case of HHG, the returning electron recombines with the parent ion, emitting a high-energy photon in the form of harmonic radiation. Alternatively, if the electron possesses sufficient kinetic energy upon return, it can liberate the second electron from the ion by, e.g., impact ionization or excitation with subsequent ionization by the laser field (see Ref.~\cite{FariaRev2011}). These processes are captured by the three-step model that advances the two-step model by incorporating the interaction of the photoelectron with the parent ion as its third and final step, see Fig.~\ref{three_step}. Therefore, the three-step model provides a basis for understanding of rescattering-induced strong-field processes. Indeed, the three-step model has successfully explained a wide range of features observed in high-order ATI, HHG, and NSDI. Well-known examples include the cutoff energies in high-order ATI spectra \cite{Paulus1994} and HHG \cite{Krause1992,Lewenstein1994}, the maximum emission angles in the photoelectron angular distributions \cite{Paulus1994b}, and the characteristic momenta of recoil ions observed in NSDI \cite{Faria2003,Milosevic2003}. 

Both the two-step \cite{Linden1988,Gallagher1988,Corkum1989} and three-step \cite{Corkum1993} models neglect the influence of the ionic potential on electron motion.
Accounting for the Coulomb interaction by including the ionic force in the classical equations of motion has led to significant refinements. This extension has made it possible to reveal the Coulomb focusing effect \cite{BrabecIvanov1996}, investigate the Coulomb-induced cusp in photoelectron angular distributions \cite{Dimitriou2004}, and study low-energy structures emerging in electron energy spectra and momentum distributions produced by strong mid-infrared laser pulses \cite{Quan2009,Liu2010,Yan2010,Kastner2012,Lemell2012,Lemell2013,Wolter2014,Becker2014,Dimitrovski2015}. The latter phenomenon, often referred to as the ionization surprise, was first observed experimentally in Ref.~\cite{Blaga2009}. The extended models have facilitated studies of nonadiabatic effects in ionization by intense laser pulses, see, e.g., \cite{Boge2013,Hofmann2014,Geng2014}. 

A second important advantage of trajectory-based models is their generally lower computational cost compared to direct numerical solutions of the TDSE - although this is not necessarily always the case. For certain strong-field processes, semiclassical methods are currently the only feasible theoretical approach. A prominent example is NSDI in atoms subjected to elliptically \cite{Shvetsov2008,XuWang2009,Hao2009} and circularly polarized fields \cite{Uzer2010}, as well as this process in molecular systems induced by linearly polarized fields (see, e.g., Ref.~\cite{Agapi2011}). It should be stressed that these cases are beyond the reach of fully quantum mechanical simulations. Therefore, further development, refinement, and application of trajectory-based models to strong-field phenomena are of great importance. 

In this paper, we review the semiclassical models aimed at the description of ATI, HHG, NSDI, and frustrated tunneling ionization. The latter term is used for the production of a substantial number of neutral atoms in the tunneling regime (i.e., for ionization by a strong and low-frequency laser pulse), see Ref.~\cite{Eichmann2008}.  First, we discuss formulations of the trajectory-based models and their practical implementations. We then focus on some of the applications of these models to the main strong-field phenomena, including some of recent ones. Nevertheless, the scope of the article does not allow us to deal with several interesting applications of semiclassical models. This applies to the semiclassical modelling of ATI in molecules (see, e.g., Refs.~\cite{Liu2016,Liu2017,Guo2022}) and studies of the multielectron polarization-induced effects in the ATI process in atoms and molecules (see Refs.~\cite{Keller2012,Shvetsov2012,Dimitrovski2014,Dimitrovski2015,Kang2018e,Shvetsov2018b} and references therein). We also only briefly discuss the semiclassical models aimed at description of interference effects in ATI. For recent review of one such models, namely, the semiclassical two-step model (SCTS), see Ref.~\cite{Shvetsov2021}.   

The paper is organized as follows. In Sec.~\ref{secin} we discuss the formulations of trajectory-based models in strong-field physics, focusing on the choice of initial conditions for classical trajectories. In Sec.~\ref{ati}, we discuss the application of the semiclassical models to ATI and high-order ATI. In Sec.~\ref{fti}, we review the semiclassical simulations of the FTI effect. Applications of trajectory-based models to the generation of high-order harmonics of the incident field are discussed in Sec.~\ref{hhg}. In Sec.~\ref{nsdi}, we briefly review some of the semiclassical simulations of the NSDI process.  The conclusions of the paper are given in Sec.~\ref{conc}. 

\section{Trajectory-based models: Theoretical background}
\label{secin}

In semiclassical models, the motion of an electron is determined by integrating Newton's equation:
\begin{equation}
\label{newton}
\frac{d^{2}\mathbf{r}}{dt^2}=-\mathbf{F}\left(t\right)-\nabla V\left(\mathbf{r},t\right),
\end{equation}
In this notation, $\mathbf{r}$ denotes the electron position, while $V\left(\mathbf{r},t\right)$ represents the parent ion potential. For simplicity, we assume that the laser field is linearly polarized along the $z$-axis, $\mathbf{F}\left(t\right)=F_z\left(t\right)\mathbf{e}_{z}$, and we treat the tunneling problem in the quasistatic approximation (see, e.g., Ref~\cite{Dau3}).  In order to integrate Eq.~(\ref{newton}), we need to know the initial conditions: the starting point and the initial velocity of the electron. 

The choice of initial conditions is a delicate aspect of semiclassical models. The initial conditions serve to bridge the classical trajectory picture with the underlying quantum description, making their selection nontrivial. In many trajectory-based models, it is assumed that the electron starts with zero initial velocity along the laser field $v_{0,z}=0$, but it can have a nonzero initial velocity $v_{0,\perp}$ in the perpendicular direction. 

A basic estimate for the tunnel exit point $z_e$ is obtained by neglecting the ionic potential, assuming a triangular barrier defined by the ground-state energy and the laser field:
\begin{equation}
\label{triangle}
r_e\left(t_0\right)=\frac{I_p}{F\left(t_0\right)}.
\end{equation}
Here $I_p$ is the ionization potential, $r_e\left(t_0\right)=\left|z_e\left(t_0\right)\right|$, and the sign of the exit point is determined by the condition that the electron tunnels through the potential barrier in the direction opposite to the field at the ionization time $t_0$. 

Another approach to determine $z_e$ is to consider a one-dimensional (1D) tunneling problem in the direction of the laser field. More specifically, the exit point is found from the equation:
\begin{equation}
\label{fdm}
V\left(z_e\right)+F\left(t_0\right)=-I_{p},
\end{equation}
i.e., assuming a cut of the potential barrier formed by the ionic potential and the laser field along the field direction. This approach is called the field direction model (FDM) \cite{Shvetsov2012}. It was used in many studies, see, e.g, Refs.~\cite{BrabecIvanov1996,Dimitriou2004,Lemell2012}. 

When the Coulomb potential of the parent ion is included, the tunnel exit point can be obtained via separation in parabolic coordinates. We define the parabolic coordinates as $\xi=r+z$, $\eta=r-z$, and $\phi=\arctan\left(y/x\right)$. The tunnel exit point $z_e=-\eta_{e}/2$ is then determined by the solution of the equation:
\begin{equation}
\label{tunex}
-\frac{\beta_{2}\left(F\right)}{2\eta}+\frac{m^2-1}{8\eta^2}-\frac{F\eta}{8}=-\frac{I_{p}\left(F\right)}{4},
\end{equation}
where $I_p\left(F\right)$ is the (Stark-shifted) ionization potential, $m$ is the magnetic quantum number of the initial state, and the separation constant $\beta_2\left(F\right)$ is calculated as follows:
\begin{equation}
\label{beta}
\beta_{2}\left(F\right)=Z-\left(1+\left|m\right|\right)\frac{\sqrt{2I_{p}\left(F\right)}}{2}.
\end{equation}
The ionization potential in Eq.~(\ref{tunex}) is generally defined as
\begin{equation}
\label{yp}
I_{p}\left(F\right)=I_p\left(0\right)+\left(\mathbf{\mu}_{N}-\mathbf{\mu}_{I}\right)\cdot\mathbf{F}+\frac{1}{2}\left(\alpha_N-\alpha_I\right)\mathbf{F}^{2},
\end{equation}
where $I_p\left(0\right)$ is the field-free ionization potential in the absence of the field, $\alpha_{N}$ is the static polarizability of an atom (molecule), and $\mathbf{\mu}_{N}$ and $\mathbf{\mu}_{I}$ are the dipole moments of an atom (molecule) and of the corresponding ion, respectively. 

An effective potential including the laser field, Coulomb potential, and ionic core polarization has been introduced in Refs.~\cite{Brabec2005,Zhao2007,Dimitrovski2010} by using the adiabatic approximation:
\begin{equation}
\label{ME_pot}
V\left(\mathbf{r},t\right)=-\frac{Z}{r}-\frac{\alpha_{I}\mathbf{F}\left(t\right)\cdot\mathbf{r}}{r^3},
\end{equation}
where $\alpha_{I}$ is the static polarizability of the ion, and the multielectron effect is accounted for through the induced dipole potential $\left[\frac{\alpha_{I}\mathbf{F}\cdot\mathbf{r}}{r^3}\right]$. The approximate separation of the tunneling problem for this potential and accounting for the Stark shift of the ionization potential is possible in parabolic coordinates \cite{Pfeiffer2012}. In this case, the tunneling coordinate $\eta_{e}$ is determined from the following equation:
\begin{equation}
\label{tunex2}
-\frac{\beta_{2}\left(F\right)}{2\eta}+\frac{m^2-1}{8\eta^2}-\frac{F\eta}{8}+\frac{\alpha_{I}F}{\eta^2}=-\frac{I_{p}\left(F\right)}{4},
\end{equation}
The approximate separation procedure leads to a certain geometry of the tunneling process, and the corresponding physical picture is referred to as tunnel ionization in parabolic coordinates with induced dipole and Stark shift (TIPIS). The semiclassical model based on the TIPIS approach has been extensively used for studies of multielectron polarization effects.  

A common approach assigns ionization times and initial transverse velocities according to the static tunneling rate (see, e.g., Ref.~\cite{Dau3}):
\begin{equation}
\label{tunrate}
w\left(t_{0},v_{0, \perp}\right)\sim\exp\left(-\frac{2\kappa^3}{3F\left(t_0\right)}\right)\exp\left(-\frac{\kappa v_{0,\perp}^{2}}{F\left(t_0\right)}\right),
\end{equation} 
with $\kappa=\sqrt{2I_{p}\left(F\right)}$. 

However, this formula produces too few trajectories with large transverse momenta. As a result, semiclassical photoelectron momentum distributions are narrower compared to the distributions obtained from the solution of the TDSE or, as a consequence of this fact, the semiclassical electron energy spectra fall off too rapidly (see, e.g., Ref.~\cite{Shvetsov2021}). The use of SFA formulas for the distribution of initial conditions appears to be an appealing alternative to the tunnelling formula (\ref{tunrate}). This idea dates back to the investigations of the instantaneous ionization yield \cite{YudinIvanov2001,Bondar2008}, and it has been used in many investigations employing semiclassical models (see, e.g., Refs.~\cite{Yan2010,Boge2013,Hofmann2014,Geng2014}). However, it should be stressed that the validity of the SFA expressions as initial conditions for classical trajectories has not been systematically investigated so far. Such investigations are necessary since results of semiclassical simulations are obviously sensitive to the distribution of the initial conditions for classical trajectories. 

Modern semiclassical models apply various sampling strategies to distribute initial conditions and/or operate with classical trajectories in different ways. In order to illustrate these strategies, here we discuss semiclassical simulations of the ATI process. The simplest approach is to launch and propagate a large ensemble of classical trajectories. After the final momenta of all the trajectories are found, they can be distributed over the cells in momentum space, and thus the photoelectron momentum distribution is calculated. In strong-field physics, this approach is often referred to as the ``shooting method", although it has nothing to do with a method for solving boundary value problems. Different sampling strategies can be used to implement the ``shooting method." In the simplest approach, the initial conditions for classical trajectories are chosen randomly or from a uniform grid. In this case, the electron momentum distribution is obtained as: 
\begin{equation}
\label{stad_noint}
R\left(\mathbf{k}\right)=\sum_{j=1}^{n_p}w\left(t_0^{j},v_0^{j}\right),
\end{equation}
All $n_p$ trajectories ending up at the given cell that corresponds to the final momentum $\vec{k}$ contribute to the sum in Eq.~(\ref{stad_noint}).
     
In the importance sampling approach, the weight of each trajectory is considered already at the sampling stage, i.e., before the time evolution of the trajectory is calculated. This is achieved by distributing the initial conditions in accordance with the tunneling rate [e.g., Eq.~(\ref{tunrate}) or the corresponding SFA formulas]. Within this approach, the ionization probability $R\left(\mathbf{k}\right)$ is found as a number of trajectories leading to the corresponding cell of the momentum space. It should be noted that more advanced sampling strategies have recently been developed, see, e.g., Ref.~\cite{Faria2023,Madsen2024}. 

An alternative to the ``shooting method" is provided by an approach based on solving the so-called ``inverse problem", i.e., finding all trajectories leading to a given asymptotic momentum. The solution of the inverse problem allows for the avoidance of the need to calculate a large ensemble of classical trajectories. Furthermore, it controls the cusps and caustics that often emerge in classical simulations more efficiently than the ``shooting method". However, the solution of the inverse problem is often very difficult. It is also clear that semiclassical models employing this approach are less versatile.  

Regardless of the approach used, one has to be able to find the final electron momentum. After the end of the laser pulse at $t=t_{f}$ the electron moves only in the potential of the ion. At large distances, this potential can be treated as the Coulomb one.  If the electron energy $E$ at the end of the laser pulse $t_{f}$ is negative $E<0$, the electron is on the elliptic orbit, and therefore, it should be considered as captured into a Rydberg state of an ion. This process of trapping initially freed electrons into bound states is often referred to as frustrated tunneling ionization (FTI). In contrast to the captured electrons, the ionized electrons have positive energies and move along hyperbolic orbits. The final (asymptotic) momentum of an ionized electron that is measured by a detector is determined by the electron position and momentum at the end of the pulse. More specifically, the following formula can be used:
\begin{equation}
\label{mominf}
\mathbf{k}=k\frac{k\left(\mathbf{L}\times\mathbf{a}\right)-\mathbf{a}}{1+k^2L^2},
\end{equation}
see Refs.~\cite{Shvetsov2009,Shvetsov2012}. Here $\mathbf{L}=\mathbf{r}\left(t_f\right)\times\mathbf{p}\left(t_f\right)$ and $\mathbf{a}=\mathbf{p}\left(t_f\right)\times\mathbf{L}-Z\mathbf{r}\left(t_f\right)/r\left(t_f\right)$ are the electron angular momentum and the Runge-Lenz vector at the end of the pulse, respectively. The absolute value of the momentum $k$ can be found from energy conservation:
\begin{equation}
\label{enrg}
\frac{k^2}{2}=\frac{\vec{p}^{~2}\left(t_f\right)}{2}-\frac{Z}{r\left(t_f\right)}.
\end{equation}

It goes without saying that the described method of distributing the initial conditions and its variations are not the only ways to set up a semiclassical model in strong-field physics. An alternative method, which was particularly popular about two decades ago, is to consider a bound electron moving classically along an ellipse. This method that is often referred to as the classical trajectory Monte-Carlo (CTMC) method was established and developed in Refs.~\cite{Abrines1966, Cohen1982}. It has been applied for simulations of atomic and molecular collisions, e.g., calculations of cross sections of impact-ionization and charge-transfer. Within this approach, the initial state of an electron is specified by its binding energy and five additional parameters randomly distributed in certain ranges. These parameters are the eccentricity of the orbit, a parameter proportional to the time, and three Euler angles \cite{Cohen1982}. For random distribution of these parameters all phases of the electron motion along the Kepler ellipse have equal probabilities. Therefore, the initial coordinates and velocities are distributed in accord with the microcanonical (uniform) distribution. 

The CTMC-T (classical trajectory Monte Carlo with tunneling) method explicitly includes tunneling events \cite{Cohen2001,Cohen2003}. In the CTMC-T method, the electron can tunnel through the potential barrier if it reaches the outer turning point. The latter is specified by the two following conditions: $p_z=0$ and $zF\left(t\right)<0$. The corresponding tunneling probability is calculated in accord with the WKB approximation (see Refs.~\cite{Kemble1935,Child1967,Connor1968}): 
\begin{equation}
\label{wkb}
w=\exp\left[-2\sqrt{2}\int_{z_{\mathrm{in}}}^{z_{\mathrm{out}}}\sqrt{V\left(z\right)-V\left(z_{\mathrm{in}}\right)}dz\right].
\end{equation}
It is assumed that the electron tunnels instantaneously from the turning point to the other side of the potential barrier, being on the same energy manifold \cite{Dimitriou2004}. It should be noted that formula (\ref{wkb}) is valid for 1D systems, and therefore an appropriate tunneling path should be specified. Usually, $z_{\mathrm{in}}$ and $z_{\mathrm{out}}$ are chosen to be the two roots of the equation
\begin{equation}
\label{inout}
V\left(z\right)=-\frac{Z}{z}+zF\left(t\right)=-\frac{Z}{z_{\mathrm{in}}}+z^{\mathrm{in}}F\left(t\right),
\end{equation}
and $\left|z^{\mathrm{out}}\right|>\left|z^{\mathrm{in}}\right|$. For simplicity, the Coulomb potential is implied in Eq.~(\ref{inout}). The CTMC-T model has been successfully applied to studies of strong-field phenomena, see, e.g., Refs.~\cite{Cohen2001,Dimitriou2004,Fu2001,ChenNam2002}. The CTMC approach is widely used in studies of NSDI in order to model the state of the second bound electron, see Sec.~\ref{nsdi}. 

\section{Semiclassical simulations of above-threshold ionization}
\label{ati}

Until roughly a decade ago, semiclassical models could not reproduce quantum interference phenomena. Specifically, purely classical post-ionization dynamics failed to capture interference structures in electron momentum distributions. This deficiency has been overcome with the emergence of the trajectory-based Coulomb SFA (TCSFA) \cite{Yan2010,YanBauer2012}, the quantum trajectory Monte Carlo model (QTMC) \cite{Li2014}, the semiclassical two-step model (SCTS) \cite{Shvetsov2016}, and the Coulomb quantum orbit strong-field approximation (CQSFA) \cite{Faria2017a,Faria2017b,Faria2018b,Faria2018a,Faria2018c} (the foundations of the CQSFA method are presented in Ref.~\cite{Faria2015}). In these approaches, each trajectory carries a phase, and the coherent sum of trajectories terminating at the same final momentum gives rise to interference patterns.

The TCSFA extends the Coulomb-corrected strong-field approximation (CCSFA), see Refs.~\cite{Popruzhenko2008a,Popruzhenko2008b}. Unlike CCSFA, where the Coulomb potential is treated perturbatively, the TCSFA accounts for both laser and atomic potentials on an equal basis. However, the TCSFA uses the first-order semiclassical perturbation theory (perturbation theory in action) \cite{PPT3} to account for the Coulomb potential in an interference phase assigned to every trajectory. The same first-order perturbation theory in action was also used in the QTMC model. The SCTS and the CQSFA account for the Coulomb potential beyond the semiclassical perturbation theory. While the SCTS model propagates large ensembles of classical trajectories, the CQSFA approach is based on the solution of the inverse problem. Since the TCSFA and CQSFA methods, although using classical trajectories, are derived from SFA and are closely related to this approach, we do not discuss them here. Undoubtedly, recent developments of both these methods deserve a separate thorough review article. In what follows, we consider the QTMC and the SCTS models. 

It is assumed in the QTMC model that the electron moves in the combined laser and Coulomb force fields. In this model, the phase associated with each trajectory reads as
\begin{equation}
\label{phas_qtmc}
\Phi^{\textrm{QTMC}}\left(t_{0},\mathbf{v}_0\right)= - \mathbf{v}_0\cdot\mathbf{r}(t_0) + I_{p}t_{0} - \int_{t_0}^\infty dt\, \left\lbrace\frac{\mathbf{p}^2(t)}{2}-\frac{Z}{r\left(t\right)}\right\rbrace.
\end{equation}
The phase of the SCTS model is calculated using the matrix element of the semiclassical propagator $U_{\textrm{SC}}\left(t_{2},t_{1}\right)$ between the initial state at time $t_1$ and the final state at time $t_2$ (see Refs.~\cite{Miller1974,WalserBrabec2003,Spanner2003} and Refs.~\cite{TannorBook2007,GrossmannBook2008} for a textbook treatment). As a result, each trajectory has the phase:
\begin{equation}
\label{phas_scts}
\Phi^{SCTS}\left(t_{0},\mathbf{v}_0\right)= -\mathbf{v}_0\cdot\mathbf{r}(t_0) + I_{p}t_{0} - \int_{t_0}^\infty dt\, \left\lbrace\dot{\mathbf{p}}(t)\cdot\mathbf{r}(t)+H[\mathbf{r}\left(t\right),\mathbf{p}(t)]\right\rbrace,  
\end{equation}
Here, $\mathbf{p}_1$ and $\mathbf{p}_2$ are electron momenta at times $t_1$ and $t_2$, respectively, $\mathbf{r}_1$ is the position at the time $t_1$, and $H\left[\mathbf{r}\left(t\right),\mathbf{p}(t)\right]$ is the classical Hamiltonian as a function of the canonical coordinate $\mathbf{r}$ and momentum $\mathbf{p}$. For an electron released at time $t_0$ and moving in laser and Coulomb fields, the phase (\ref{phas_scts}) reads as:
\begin{equation}
\label{phas_scts_coul}
\Phi^{\textrm{SCTS}}\left(t_{0},\mathbf{v}_0\right)= - \mathbf{v}_0\cdot\mathbf{r}(t_0) + I_{p}t_{0} - \int_{t_0}^\infty dt\, \left\lbrace\frac{\mathbf{p}^2(t)}{2}-\frac{2Z}{r\left(t\right)}\right\rbrace.
\end{equation}
The SCTS phase differs from that of QTMC [Eq.~(\ref{phas_qtmc})]  because it incorporates the Coulomb potential beyond the perturbative approximation used in QTMC.  Indeed, the Coulomb term enters with double weight in Eq.~(\ref{phas_scts}).

\begin{figure}[h]
\centering
\includegraphics[width=0.9\textwidth]{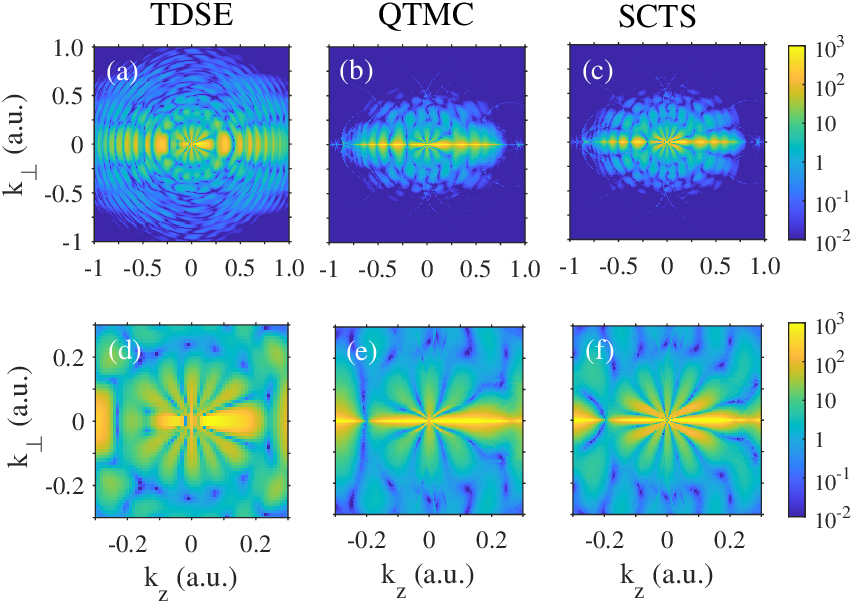}
\caption{Two-dimensional momentum distributions for the H atom ionized by a laser pulse with a duration of $8$ cycles, wavelength of $800$ nm, and peak intensity of $0.9\times10^{14}$ W/cm$^2$ obtained from (a,d) solution of the TDSE, (b,e) QTMC, and (c,f) the SCTS model. Panels (d), (e), and (f) show the maginifications for $\left|k_z\right|$, $\left|k_z\right|<0.3$~a.u. of the distributions presented in (a), (b), and (c), respectively. The distributions are normalized to the total ionization yield. A logarithmic color scale in arbitrary units is used. The laser field is linearly polarized along the $z$-axis.}
\label{SCTS_Large}
\end{figure}

The SCTS model was compared with the direct numerical solution of the TDSE and predictions of the QTMC model \cite{Shvetsov2016,Shvetsov2021}. The electron momentum distributions calculated within all three approaches for the ionization of the hydrogen atom are shown in Fig.~\ref{SCTS_Large}. Both semiclassical approaches capture the principal features of the TDSE distributions, such as elongation along the polarization axis and ATI rings, although they tend to underestimate the spread along the polarization direction due to the choice of $v_z=0$ (see Ref.~\cite{Shvetsov2016}). 

However, remarkable deviations are found in the low-energy part of the distributions. It is seen that for $\left|k\right|<0.3$~a.u., i.e., photoelectron energies below $U_{p}=0.2$~a.u., the 2D momentum distributions have fanlike interference structures. These interference structures are similar to those characteristic of Ramsauer-Townsend diffraction oscillations \cite{Arbo2006a,Arbo2006b,Gopal2009,Arbo2008}. The SCTS model reproduces the nodal pattern predicted by the TDSE, whereas the QTMC predicts fewer nodal lines, see Fig.~\ref{SCTS_Large}~(d)-(f). This is due to the underestimation of the Coulomb effect in the phase of the QTMC model, see Eq.~(\ref{phas_qtmc}).

\begin{figure}[h]
\centering
\includegraphics[width=0.5\textwidth]{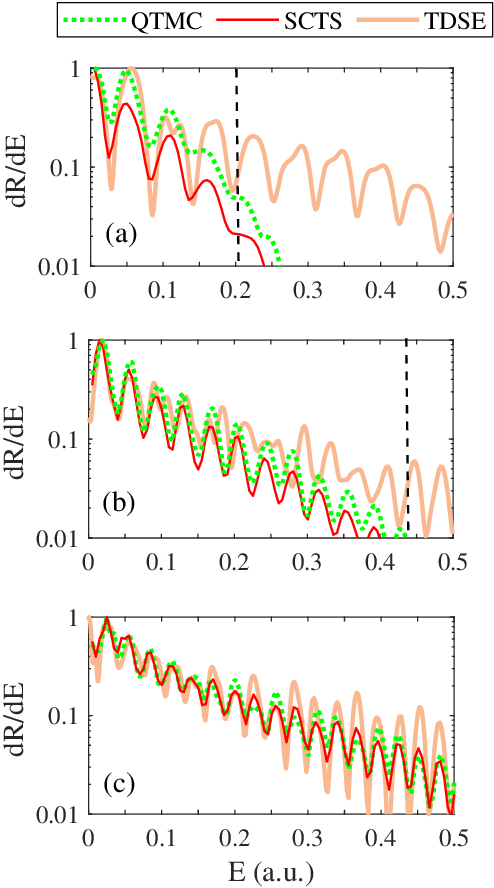}
\caption{Electron energy spectra calculated for the H atom ionized by a laser pulse with an intensity of $0.9\times10^{14}$ W/cm$^2$ and duration of $8$ cycles using numerical solution of the TDSE (thick light orange curve), the QTMC model (dotted green curve), and the SCTS model (solid red curve). Panels (a), (b), and (c) correspond to the wavelengths equal to 800 nm, 1000 nm, and 1200 nm, respectively. The energy spectra are normalized to the peak value. The dashed vertical lines show the energy equal to ponderomotive potential in panels (a) and (b).}\label{spectra}
\end{figure}

The comparison of the energy spectra and angular distributions of the photoelectrons has shown that both semiclassical models qualitatively reproduce the ATI peaks, see Fig.~\ref{spectra} and  Ref.~\cite{Shvetsov2016} for details. However, quantitative agreement between the interference structure predicted by the TDSE and the results of the semiclassical simulations can be achieved only for the first few peaks. The reason for this is that the initial conditions used in both models result in too few trajectories with large momenta along the polarization direction \cite{Shvetsov2016}. The insufficient number of trajectories with large longitudinal momenta also causes the overly rapid fall of the electron energy spectra calculated using the QTMS and the SCTS. The main reason behind the lack of full quantitative agreement between the STCS and the TDSE are inaccuracies in the choice of initial conditions for classical trajectories, i.e., the description of the tunneling step.
For comparisons, TDSE results are taken as reference, though extracting momentum distributions from the wavefunction can be nontrivial and may introduce errors, see Ref.~\cite{Milosevic2020} for details. 

Recently, an efficient extension of the SCTS model has been proposed \cite{Brennecke2020}. The effect of the preexponential factor of the semiclassical matrix element that was neglected in the original version of the SCTS was studied in Ref.~\cite{Brennecke2020}. This preexponential factor is a complex number. Its modulus describes the mapping from the space of the space of initial conditions to the electron momentum components, and therefore it affects the weights of trajectories. The phase of the preexponential factor (Maslov phase) can be treated as the case of Gouy's phase anomaly. 
It is shown that taking this phase into account modifies interference patterns of electron momentum distributions. The SCTS model was implemented by using the SFA and the saddle-point method 
It should also be noted that this study relies on the solution of the inverse problem, and a novel way of solving this problem employing a modern clustering algorithm was proposed in Ref.~\cite{Brennecke2020}. The modified SCTS model demonstrates a quantitative agreement with the direct numerical solution of the TDSE. 

It is also worth noting the semiclassical two-step model with quantum input (SCTSQI) \cite{Shvetsov2019}. The SCTSQI is a mixed quantum-classical approach to ionization by a strong laser pulse. 
In this model the initial conditions for classical trajectories are obtained from the exact quantum dynamics. More specifically, in the STCQI model, the TDSE is solved in a relatively small area of space with absorbing boundary conditions, and the Gabor transform is applied to the part of the wave function that is absorbed at each time step. This absorbed part is then transformed into an ensemble of classical trajectories. Quantum amplitudes assigned to classical trajectories are determined by the Gabor transform. It is shown that the SCTSQI corrects the inaccuracies of the SCTS in the description of the ionization step. Indeed, quantitative agreement with the solution of the TDSE has been achieved in Ref.~\cite{Shvetsov2019}. The SCTSQI model has so far been implemented for a 1D model atom only. Although this model can be straightforwardly generalized to a 3D atomic or molecular system, further work is needed to achieve computational efficiency of the SCTSQI.     
 
Currently, there is only a relatively small number of studies in which the \textit{entire} high-energy above-threshold ionization (ATI) plateau up to an energy of 10U$_{p}$ has been calculated by launching an ensemble of classical trajectories, see, e.g., Refs. \cite{LiuZhao2014,Wu2020}, with the study of Ref.~\cite{Wu2020} pertaining to the ionization of a nanotip. Both these calculations are performed without considering quantum interference. Even though the momentum distributions of photoelectrons in the plateau region can be obtained by solving the TDSE or applying the quantitative rescattering theory, calculating these distributions using trajectory-based models is of interest. It should be noted that we are not aware of any work in which the momentum distributions of electrons in the plateau, their energy spectra, or angular distributions are calculated using any semiclassical model that accounts for the ionic potential and \textit{interference}. This seems to be related to the fact that the bin size in momentum space, necessary for reliably describing interference oscillations, decreases with the increase in photoelectron energy and becomes small near the plateau cutoff. Indeed, the phase of semiclassical models contains a term proportional to the final electron energy: $\Phi \sim p^2/2$.  Thus, when the final momentum of the photoelectron changes by $\Delta p$, the phase changes by $p\Delta p$.  Comparing this change in phase with a phase difference of $\pi$ yields an estimate of the bin size in momentum space $\Delta p\sim \pi/p$. Obviously, decreasing the bin sizes requires a greater number of trajectories falling into it, i.e., a greater number of trajectories in the ensemble. This increases the computational costs of semiclassical modeling considerably. A possible solution would be the use of bins of variable size in momentum space and/or more sophisticated strategies for sampling initial conditions (e.g., a denser grid in regions of the $\left(t_0,v_0\right)$-plane that predominantly yield rescattered trajectories, followed by multiplying the corresponding contributions by certain weights). Thus, achieving a reliable semiclassical description of the high-energy ATI plateau that includes both interference and the ionic potential remains an open challenge.

\section{Trajectory-based simulations of frustrated tunneling ionization}
\label{fti}

In strong-field physics, stabilization describes the reduction of ionization yield at laser field strengths above a certain threshold. This phenomenon attracted a lot of attention in the 1980s and 1990s (see Refs.~\cite{FedorovBook1997,DeloneBook2000} for reviews). Stabilization has been attributed to two mechanisms: adiabatic (Kramers-Henneberger) stabilization and interference stabilization. Adiabatic stabilization corresponds to the situation where the laser frequency exceeds the ionization potential, $\omega>I_{p}$, and the states of the discrete spectrum and states of the continuum are coupled with a single photon energy. The mechanism of interference stabilization is different: it is due to multiple Raman-type transitions from the Rydberg states to the common continuum. In interference stabilization, ionization is suppressed through destructive interference between transition amplitudes connecting coherently populated Rydberg states to the continuum.

The study of Ref.~\cite{Eichmann2008} that introduced the term frustrated tunneling ionization (FTI) stimulated further investigations of the production of neutral atoms in ionization by strong laser fields. In Ref.~\cite{Eichmann2008} the yield of neutral excited He atoms produced in ionization of a gaseous target by 30 fs laser pulses with a wavelength of 800 nm and intensities up to $10^{15}$~W/cm$^2$ was experimentally studied. For moderate laser intensities, the yield of excited neutral atoms was around 20\% of the ion yield, decreasing to ~10\% at higher intensities. These findings were compared with TDSE calculations and semiclassical simulations. The semiclassical model \cite{Eichmann2008} includes both laser and Coulomb interactions, and electrons remaining with negative energies after the pulse are considered captured into bound states. It was shown that the semiclassical distribution over the effective principal quantum number is in quantitative agreement with the TDSE result. 

The theoretical study of Ref.~\cite{Shvetsov2009} also applying a semiclassical model, focuses on the mechanism underlying the production of neutral excited atoms and the scaling of the yield of these atoms with the laser-atom parameters. Furthermore, the effect of FTI on the momentum distributions of ionized electrons was also studied \cite{Shvetsov2009}. The analysis of electron trajectories has revealed that only those electrons are captured into bound states that (i) have moderate initial transverse velocities at the ionization time, and (ii) avoid hard collisions with the parent ion (see Ref.~\cite{Shvetsov2009} for details). This analysis has made it possible to derive the following analytical estimate for the ratio of the number of neutral excited atoms $N^{*}$ and singly charged ions $N^{+}$:
\begin{equation}
\frac{N^{*}}{N^{+}}\sim\frac{\omega}{F^{3/2}\tau_{L}^{2/3}}\left(1-2\frac{F}{\left(2I_{p}\right)^2}\right)^{-1},
\end{equation}
where $\tau_{L}$ is the duration of the laser pulse. This scaling with the laser-atom parameters is in a good agreement with semiclassical simulations. In Ref.~\cite{Ortmann2018}, a semiclassical simulation was used to obtain scalings of the capture into Rydberg states, accounting for nonadiabatic effects. Recently, a semiclassical model and the analysis of classical trajectories were also applied to control the principal quantum numbers of neutral excited atoms \cite{Ortmann2021}.

\begin{figure}[h]
\centering
\includegraphics[width=0.9\textwidth]{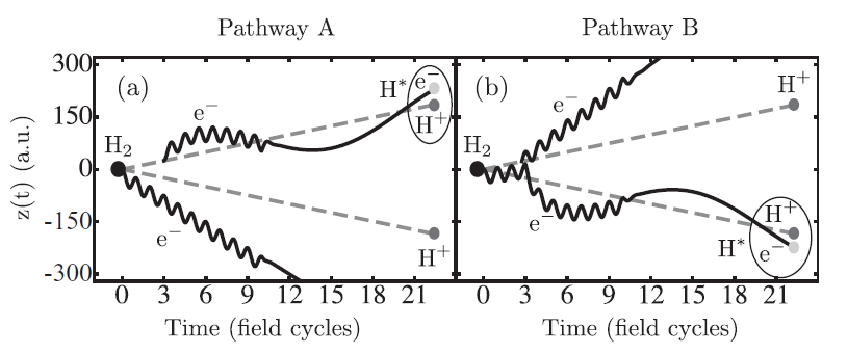}
\caption{Scheme of the two routes leading to the formation neutral excited H atoms: Pathway A (a) and pathway B (b). Black curves and gray dashed lines show the time-dependent position of electrons and ions, respectively. From \href{https://doi.org/10.1103/PhysRevA.85.011402}{A.~Emmanouilidou, 2012}}\label{pathways}
\end{figure}

Formation of highly excited neutral H atoms in the interaction of hydrogen molecules with strong laser pulses is investigated in Ref.~\cite{Emma2012}.  Therefore, the products of this interaction are an excited neutral H atom, a singly charged hydrogen ion, and a liberated electron. A process of this kind may be called frustrated double ionization (FDI), since only one of the two electrons of the H$_2$ molecule eventually escapes \cite{Emma2012,Emma2014}.  In Ref.~\cite{Emma2012}, the full four-body Hamiltonian was propagated in time using a regularization of the coordinates \cite{Heggie1974} to eliminate Coulomb singularities. Importantly, the previously developed 3D semiclassical model did not account for the nuclei motion. The ionization rate of the first electron was described by the semiclassical formula of Ref.~\cite{Ye2008}. The initial vibrational state of the nuclei is taken to be the ground state of the Morse potential. The Wigner distribution of this state was applied in order to describe the initial state of the nuclei. These results demonstrate for the first time that accounting for two-electron effects is important for the understanding of all pathways leading to the formation of highly excited H atoms \cite{Emma2012}. Two such pathways were identified in Ref.~\cite{Emma2012}: the electron that is released from the molecule either escapes very quickly, or after it remains bound for several periods of the laser pulse, see Fig.~\ref{pathways}. Notably, these two pathways have distinct imprints on the momentum distributions of liberated electrons.

A computational toolkit enabling semiclassical simulations of molecules in strong laser fields was developed in Ref. ~\cite{Emma2014}. This method combines leapfrog propagation with the Bulirsch-Stoer scheme \cite{Bulirsch1966}  (see Ref.~\cite{Press} for a textbook treatment), regularizes Coulomb singularities, and provides a stable treatment for systems with large mass ratios. Furthermore, the method of Ref.~\cite{Emma2014} allows for tunneling of each of the two electrons during the time propagation, i.e., the motion of both electrons is not entirely classical. This feature is important for an accurate description of the enhanced ionization process. Indeed, in this process a double potential well is formed at a critical internuclear distance.  As a result, it is easier for an electron bound in the higher potential well to tunnel to the lower potential well and then leave the molecule in the ionization process. The developed toolkit was used to study the FDI process in an elliptically polarized laser field. It is shown that the two-electron effects disappear with increasing ellipticity \cite{Emma2014}.  

Using the toolkit \cite{Emma2014}, the FDI process in D$_3^{+}$ was studied in Ref.~\cite{Emma2016}. A D$_3^{+}$ is a two-electron triatomic molecule, and therefore, FDI for this molecule is described by the following scheme: D$_3^{+}\rightarrow$D$^{*}$+2D$^{+}+e$. It was found that the kinetic energy release of the nuclei calculated within the semiclassical method is in a good agreement with experimental results. The two pathways responsible for FDI were investigated for the D$_3^{+}$ molecule. It is shown that with increasing intensity the underlying mechanism switches from tunneling to barrier-suppression ionization. Finally, the angular distributions of the ion fragments show an asymmetry that is a signature of only one of the two pathways of the FDI process \cite{Emma2016}. This asymmetry can be observed experimentally. 

The toolkit \cite{Emma2014} turned out to be very powerful in studies of the FDI process. It was applied to the control of electron-electron correlation in FDI with two orthogonally polarized two-color laser fields \cite{Emma2017,Vila2018a}, the study of the FDI process in a heteronuclear HeH$^{+}$ molecule \cite{Vila2018b}, and enhancing FDI with no electronic correlation using counter-rotating two-color circular laser fields \cite{Emma2020}. The study of Ref.~\cite{Emma2017} focuses on the control of electron-electron correlation in FDI of the  D$_3^{+}$ molecule by using two orthogonally polarized two-color laser fields. It is shown that the control over this process can be achieved by varying the time delay between the two laser pulses: the probability of the FDI process changes sharply with this delay. A characteristic split in the momentum distribution of the liberated electron was identified. The split occurs for time delays corresponding to minima of the FDI probability. It is shown that the split is a signature of the absence of electron-electron correlation in FDI \cite{Emma2017}. The sub-cycle control of DFI of the D$_3^{+}$ molecule was achieved in Ref.~\cite{Vila2018a} using two orthogonally polarized two-color laser fields. In doing so, the FDI process triggered by a strong near-infrared laser field is probed by a weak mid-infrared field. The FDI probability and the momentum distribution of the released electron along the polarization direction of the mid-infrared field considered in conjunction show clear signatures of the sub-cycle control of the process \cite{Vila2018a}. It is shown that momentum distributions of the escaping electron have a characteristic hive-like shape. The features of this shape can be mapped to the time that one of the two electrons tunnels out of the molecule at the start of the breakup process. 

The FDI process in a heteronuclear HeH$^{+}$ molecule was studied in Ref.~\cite{Vila2018b}.  The distributions over the sum of the final nuclei kinetic energies and the principal quantum number were compared to the corresponding distributions for the H$_2$ molecule. It is found that the distribution over the sum of the final kinetic energies of the nuclei has more than one peak for HeH$^{+}$ interacting with a strong laser pulse. This feature is absent for the H$_{2}$ molecule. Similarly, the distribution over $n$ has several peaks for HeH$^{+}$, whereas only a single peak and a ``shoulder" are present in this distribution for the H$_2$ molecule. This latter feature is a clear signature of the intertwined electron-nuclear motion. A significant enhancement of the FDI process in the D3+ molecule was demonstrated in Ref.~\cite{Emma2020} with counter-rotating two-color circular (CRTC) laser fields.  A pathway of FDI that is absent for molecules interacting with strong linearly polarized laser field was identified and analyzed \cite{Emma2020}. It is found that it is this pathway, in which the first ionization step is frustrated and there is no electron-electron correlation, that is responsible for the enhancement of the process in CRTC fields.  

The computational method~\cite{Emma2014} aimed at description of the H$_2$ and D$_3^{+}$   molecule was further generalized to more complex atomic and molecular systems in Refs.~\cite{Emma2021} and \cite{Emma2022}. It should be noted that the main problem with 3D semiclassical methods aimed at the description of multielectron ionization and accounting for the Coulomb potential is unphysical autoionization. This process occurs in the following way: one of the bound electrons treated classically can undergo a close encounter with the ionic core and gain a large negative energy. The energy released in this process leads to the release of the other bound electron. Quantum mechanical approaches do not show this problem due to the lower bound of the electron energy. The unphysical autoionization in 3D trajectory-based models can be eliminated by adding a Heisenberg potential \cite{Kirschbaum1980}. The Heisenberg potential leads to adding a potential barrier that prevents electrons from a close encounter of the core. The potential barrier mimics the Heisenberg uncertainty principle. However, the momentum of a particle is no longer related to the derivative of the position vector in this approach \cite{Cohen1995,Cohen1996}. This is a consequence of adding to the Hamiltonian an extra term depending on position and momentum. 

A 3D semiclassical model of triple ionization and frustrated triple ionization of the linear HeH$_{2}^{+}$ molecule was developed in Ref.~\cite{Emma2021}. This model fully accounts for the Coulomb singularities, and it avoids unphysical autoionization by switching off the Coulomb force between two bound electrons. The Coulomb force is then switched on when the motion of one of the electrons is mainly determined by the laser field. The corresponding criteria were formulated \cite{Emma2021}. By using this model, both triple and frustrated triple ionization were investigated in Ref.~\cite{Emma2021}. In the process of frustrated triple ionization (FTI), two electrons are released from an atom or molecule while the third one remains bound, populating a Rydberg state. It is found that similar to the FDI process, two different pathways dominate in FTI. It is also more probable that the electron bound in a Rydberg state is more likely attached to He$^{2+}$  \cite{Emma2021}. It is shown that electronic correlation is weak in both triple ionization and FTI. 

The approach \cite{Emma2021} was further modified and generalized to the case of three-electron atoms in Ref.~\cite{Emma2022}. More specifically, the interaction between a pair of bound electrons is described by an effective Coulomb potential \cite{Montemayor1989}. Simultaneously, the Coulomb interaction between the electrons that are not bound is fully taken into account. Therefore, the model \cite{Emma2022} is referred to as effective Coulomb potential for bound-bound electrons (ECBB). An efficient set of criteria that are capable of determining whether an electron is bound or ``quasifree" during time propagation (i.e., on the fly) was developed in Ref.~\cite{Emma2022}. The ECBB model can be used for studies of atoms with more than three electrons. The model is formulated within the nondipole approximation, incorporating the magnetic component of the Lorentz force. The ECBB model was applied to triple and double ionization of the Ar atom by a strong few-cycle laser pulse. It is found that predictions of the model are in good agreement with experimental results for double ionization of Ar \cite{Chen2017}. 

\section{Semiclassical models and high-order harmonic generation}
\label{hhg}

Although many studies have investigated HHG spectra using semiclassical models, the number is smaller than might be anticipated, see Refs.~\cite{Chu1990,Maquet1992,Keitel1995,Duan2000,Lee2001,Botheron2009,Uzdin2010,Higuet2011,Cloux2015,Shafir2012,Soifer2010,Hostetter2010,Abanador2017,Solanpaa2014}. The earliest study of this type was published in 1990 \cite{Chu1990}, prior to the development of the three-step model. In Ref.~\cite{Chu1990} it was suggested to compute the harmonic spectrum $I\left(\omega\right)$ as the spectral density of the dipole moment $z\left(t\right)$: 
\begin{equation}
\label{hhg}
I\left(\omega\right)=\frac{1}{2\pi}\lim_{T\rightarrow 0}\frac{1}{2T}\left<\left|\int_{0}^{2\pi}z\left(t\right)\exp\left(i \omega t\right)dt\right|^2\right>,
\end{equation}
where $\left<\right>$ is the average over the ensemble of relevant classical trajectories. The hydrogen atom was considered in Ref.~\cite{Chu1990}. The necessary initial conditions for calculation of trajectories were sampled according to the microcanonical distribution. The study analyzed the underlying physical mechanisms of ionization using representative classical trajectories. As a result, the harmonic spectra were calculated only for some trajectories typical for strong-field ionization \cite{Chu1990}.

The single-trajectory analysis of the dipole moment and power spectra generated in the ionization of atomic hydrogen was also carried out in Ref.~\cite{Maquet1992}. At the same time, the HHG spectra calculated by using an ensemble of classical trajectories and Eq. ~(\ref{hhg}) were presented \cite{Maquet1992}. As in Ref.~\cite{Chu1990}, the initial conditions were sampled based on the microcanonical distribution. The parameters  used in \cite{Maquet1992} correspond to the tunneling regime: $\omega=0.086$~a.u. and $F=0.11-0.14$~a.u. (the Keldysh parameter $\gamma=0.61-0.78$). The resulting HHG spectra were found to be noisy, and the level of the noise increased with increasing intensity. The origin of this noise was studied \cite{Maquet1992}. The HHG process in the relativistic regime at the intensities exceeding $3.5\cdot10^{16}$ W/cm$^2$ was investigated in Ref.~\cite{Keitel1995}. Among other issues, the HHG spectrum in this regime was calculated using a trajectory-based model. 

The study of Ref.~\cite{Duan2000} considers ionization, dissociation, and HHG in the hydrogen molecular ion using a semiclassical model. This study proposed a method to generate initial conditions for semiclassical simulations using the Born-Oppenheimer approximation. A more powerful and sophisticated semiclassical model for HHG in hydrogen atom and the H$_{2}^{+}$ ion was developed in Ref.~\cite{Lee2001}. In this model, the electron-nuclei interaction is treated by using a regularization procedure. The procedure relies on transformation to parabolic coordinates and the usage of a fictitious time scale (see Ref.~\cite{Lee2001} for details). In order to generate the initial microcanonical distribution, action-angle variables were used. HHG spectra calculated using this model demonstrate a characteristic plateau with only odd harmonics. As in Ref.~\cite{Maquet1992}, the spectra calculated in \cite{Lee2001} become noisier at higher intensities. This was explained by the emergence of chaotic behavior in the atom (ion) exposed to the laser field. 

The HHG process in a one-dimensional hydrogen atom was analyzed in Ref.~\cite{Botheron2009}. Such an atom is described by a Hamiltonian $H=p^2/2-1/\sqrt{x^2+1}+xF\left(t\right)$, where the soft-core parameter $a$ in the potential term is used to remove the Coulomb singularity at $x=0$. A microcanonical distribution \cite{Abrines1966} was used to sample the initial conditions of classical trajectories. Unlike earlier studies, Ref.~\cite{Botheron2009} applied a discrete representation of the phase-space distribution for semiclassical simulations:
\begin{equation}
\rho\left(x,p,t\right)=\frac{1}{N}\sum_{j=1}^{N}\delta\left(x-x_j\left(t\right)\right)\delta\left(p-p_j\left(t\right)\right),
\end{equation}
where $x\left(t\right)$ and $p\left(t\right)$ are the time-dependent coordinate and momentum of the $j$-th trajectory, respectively. The ionized part of the classical distribution is extracted using the condition $E>0$.  The classical ionized density in coordinate space $\rho_{C}^{\textrm{ion}}\left(x,t\right)$ is found by integrating $\rho_{C}^{\textrm{ion}}\left(x,p,t\right)$ over $p$. In order to remove noise, this density is smoothed by a Gaussian function with the width $\Delta x$:
\begin{equation}
\label{dens_botheron}
\rho_{C}^{\textrm{ion}}\left(x,t\right)=\frac{1}{N}\sum_{j,E_{j}>0}\frac{1}{\sqrt{2\pi}\Delta x}\exp\left(-\frac{\left[x-x_{j}\left(t\right)\right]^2}{2\Delta x^2}\right)
\end{equation}
The density of Eq.~(\ref{dens_botheron}) is used to calculate the dipole $d_{C}\left(t\right)=\int \rho_{C}\left(t\right)xdx$. The HHG spectrum is obtained as a Fourier transform of the dipole acceleration:
\begin{equation}
I\left(\omega\right)=\left|\frac{1}{\sqrt{2\pi}}\int_{0}^{T}\ddot d_{C}\left(t\right)\exp\left(-i \omega t\right)dt\right|^2
\end{equation}
In contrast to other studies, in Ref.~\cite{Botheron2009} semiclassical HHG spectra were compared to the spectra obtained from the direct numerical solution of the TDSE. It was shown that the semiclassical model correctly reproduces the first odd harmonics of the quantum spectrum. However, the plateau of high-order harmonics and its cutoff are not seen in the semiclassical spectrum. The latter decreases continuously with increasing energy of the emitted photon \cite{Botheron2009}.

The study of Ref.~\cite{Uzdin2010} focuses on the difference between harmonic spectra calculated using single electron trajectories and the spectra obtained by averaging over an ensemble of classical trajectories.
Single-trajectory spectra display features that are not observed in experiments or quantum calculations; averaging over an ensemble removes these unphysical artifacts, recovering the characteristic HHG features, including the plateau, cutoff, and the presense of the odd harmonics only. This ``ensemble effect" was isolated and thoroughly studied \cite{Uzdin2010}. The study in \cite{Uzdin2010} has found the smallest and simplest subensemble of an ensemble of classical trajectories that is required in order to obtain only the physical features of the HHG process. It should be noted that only relatively low laser intensities corresponding to a Keldysh parameter of about 2.0 are considered in Ref.~\cite{Uzdin2010}. 

\begin{figure}[h]
\centering
\includegraphics[width=0.6\textwidth]{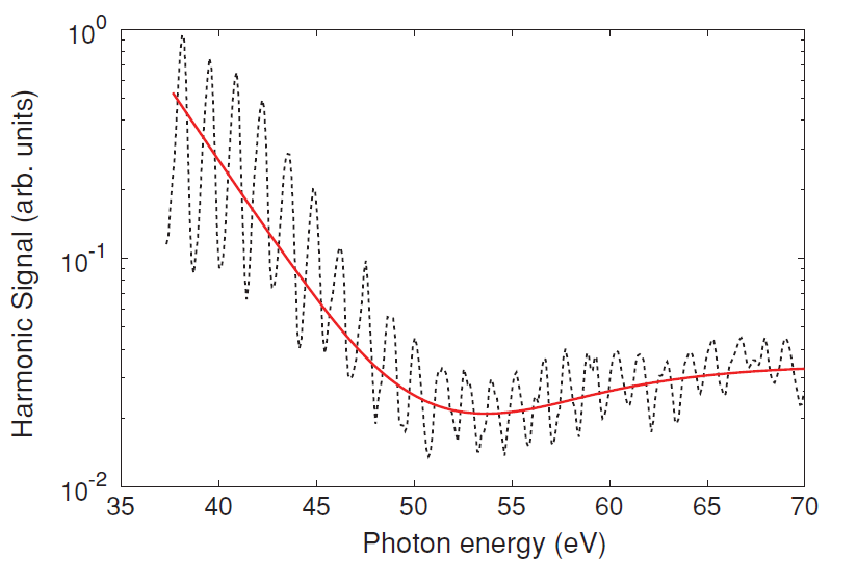}
\caption{Experimental (dashed curve) and CTMC-QUEST high-order harmonics spectra obtained for a laser pulse with a wavelength of 1380 nm and a peak intensity of $1.0\times10^{14}$ W/cm$^2$. From \href{https://doi.org/10.1103/PhysRevA.83.053401}{J.~Higuet et al.}}\label{quest}
\end{figure}

The Cooper minimum in HHG from argon atoms was investigated in Ref.~\cite{Higuet2011} both experimentally and theoretically. The theoretical approach used in \cite{Higuet2011} includes the photorecombination probability in the trajectory-based model. We note that the latter employs a combination of microcanonical distributions in order to distribute initial conditions of classical trajectories. The photorecombination probability is obtained quantum mechanically by using the Fermi's golden rule and the microreversibility principle. The resulting method is referred to as classical trajectory Monte-Carlo - quantum electron scattering techniques (CTMC-QUEST) in Ref.~\cite{Higuet2011}. It is shown that the CTMC-QUEST method accounts for the influence of the ionic potential on the wave packet of the recolliding electron, and therefore it is able to reproduce accurately the experimental spectra of high-order harmonics, see Fig.~\ref{quest}. The CTMC-QUEST method was applied for investigation of Cooper minimum in krypton \cite{Cloux2015} as well as for the study of the role of the ionic potential in the HHG process \cite{Shafir2012}.

Many semiclassical approaches calculate HHG spectra assuming that the frequency of the emitted photon is determined by the energy of an electron upon its return to the parent ion. This method was successfully applied for studies of near-threshold \cite{Soifer2010} and below-threshold harmonics \cite{Hostetter2010} (in this study, the phase accumulated by the electron wave function along the returning trajectory is also calculated), HHG in elliptically polarized fields \cite{Abanador2017}, and optimal control of the HHG process \cite{Solanpaa2014}.  

Summarizing the above, we note that electrons do not recombine in a classical system, and therefore they do not emit high-order harmonics. This leads to problems in semiclassical simulations of HHG, since the quantum recombination process must be artificially inserted in a trajectory-based model. In modern semiclassical simulations, this is done by using two different approaches. It is either assumed that  the returning electron emits a photon with energy equal to the sum of the instantaneous energy of the electron and the ionization potential \cite{Soifer2010,Hostetter2010,Abanador2017,Solanpaa2014} or a quantum recombination dipole matrix element is combined with classical density of returning electrons (CTMC-QUEST model). However, both these hybrid approaches introduce some limitations. For example, this applies to producing accurate interference patterns in HHG spectra, which is difficult to achieve using semiclassical models due to the lack of the phase information. There is a lack of  studies comparing the semiclassical simulations with the TDSE results for the generation of high-order harmonics \cite{Liu2019p}. Therefore, further developments of trajectory-based models for HHG are needed. 

\section{Semiclassical studies of nonsequential double ionization}

\subsection{Nonsequential double ionization of atoms by linearly polarized fields}
\label{nsdi}

Nonsequential double and multiple ionization has been attracting considerable attention from experimentalists and theorists for 50 years already. Indeed, the phenomenon of NSDI was discovered by V.~V.~Suran and I.~P.~Zapesochnyi in their experiment with alkaline earth atoms in 1975 \cite{Suran1975}. Seven years later, A.~L'Huillier and co-authors observed this effect for noble gas atoms \cite{Huillier1982,Huillier1983}. It should be noted that the mechanisms underlying NSDI are different for noble gas and alkaline earth atoms. In what follows, we do not consider alkaline earth atoms. NSDI in infrared fields and at higher laser intensities was first experimentally investigated in Refs.~\cite{Walker1993,Walker1994}. The interest in NSDI is caused by the fact that electron-electron correlation is a necessary prerequisite for the emergence of this phenomenon. This distinguishes NSDI from other strong-field phenomena, such as ATI, HHG, or FTI. By the present time there are already hundreds of papers exploring various aspects of this phenomenon (see, e.g., Refs.~\cite{FariaRev2011,BeckerRev2012} for reviews).

It is difficult to overestimate the role of semiclassical approaches in studies of NSDI. Suffice it to say that the breakthrough idea of the atomic antenna \cite{Kuchiev1986}, i.e., that the liberated electron absorbs energy from the laser field and gives it to its parent ion in the rescattering event, was visualized by the three-step model \cite{Corkum1993}. The advantages of semiclassical approaches are particularly important in the investigation of such a complex phenomenon as NSDI. Indeed, the solution of the TDSE for two electrons (He atom) interacting with an external laser field in three spatial dimensions is an extremely difficult problem \cite{Taylor2006}. This problem becomes prohibitively difficult for an elliptically polarized laser field and for NSDI in molecules. In both cases, semiclassical models provide a computationally feasible alternative to the solution of the TDSE. More importantly, the ability to provide an illustrative picture of the phenomena in terms of electron trajectories has allowed us to understand many aspects of NSDI. Here we briefly discuss only two remarkable examples: the study of the fingerlike structure in distribution of correlated electron momenta along the polarization direction and the investigation of NSDI below the recollision threshold.

\begin{figure}[h] 
\centering
\includegraphics[width=0.9\textwidth]{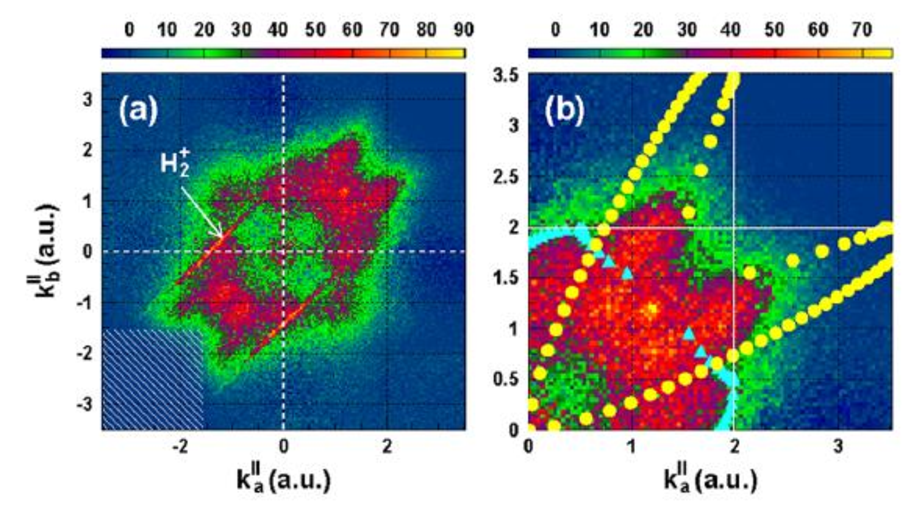}
\caption{(a) Electron momentum-momentum correlations for NSDI in helium at the intensity of $4.5\times10^{14}$ W/cm$^2$ and the wavelength of 800~nm. The first and the third quadrants correspond to both electrons emitted into the same hemisphere. (b) The first quadrant of the correlation shown in panel (a). The triangles and circles show the kinematic boundaries obtained from a simple classical model. From \href{https://doi.org/10.1103/PhysRevLett.99.263002}{Staudte et al., 2007}}\label{fingers}
\end{figure}

A breakthrough in experimental studies of NSDI was achieved with the emergence of Cold Target Recoil Ion Momentum Spectroscopy (COLTRIMS) \cite{Ullrich2003}. In combination with a high-repetition rate laser, which is required to get reliable statistics of rare events, the COLTRIMS technique has made it possible to measure differential distributions of the NSDI process, e.g., distributions of correlated electron momenta along the laser polarization direction. In 2007, two experimental groups observed a fingerlike (V-shaped) structure in the distribution of the correlated electron momenta parallel to the polarization direction (see Fig.~\ref{fingers} and Refs.~\cite{Staudte2007} and \cite{Rudenko2007}).

Although this finding is in qualitative agreement with the calculations using the S-matrix approach \cite{Faria2004a,Faria2004b,Faria2005} and the results of the direct numerical solution of the 3D TDSE, the physical origin of the fingerlike structure remained unclear. Indeed, $S$-matrix calculations employing Volkov states agree with experimental data provided electron-electron interaction is modelled by a contact potential. Simultaneously, the application of the Coulomb potential leads to deviations with the data. On the other hand, it is difficult to reveal the mechanism behind the fingerlike structure from the solution of the TDSE.

The mechanism responsible for the formation of the finger-like structure was revealed in Ref.~\cite{Liu2008} using a semiclassical model of NSDI. This model considers ionization of a helium-like atom with two active electrons. The first electron is liberated from the atom by tunneling and it starts its motion at the tunnel exit point determined by the effective potential in parabolic coordinates [see Eq.~\ref{tunex2}]. The ionization time and initial transverse velocity are distributed according to the tunneling rate (\ref{tunrate}). The initial position and momentum of the second electron follow the microcanonical distribution \cite{Abrines1966}. The evolution of the system of two electrons is described by Newton’s equations of motion:
\begin{equation}
\label{newton_nsdi}
\frac{d^2\mathbf{r}_{j}}{dt}=\mathbf{F}\left(t\right)-\nabla_{\mathbf{r}_{j}}\left(V^{j}_{ne}+V_{ee}\right)
\end{equation}
where the index $j=1,2$ denotes the two electrons, $V^{j}_{ne}=-2/{r_{i}}$ are the Coulomb interactions between the nucleus and electrons, and $V_{ee}=1/\left|\mathbf{r}_2-\mathbf{r}_1\right|$ is the Coulomb interaction between the electrons. The NSDI events were identified by using the energy criterion. According to this criterion, a double ionization event occurs when the energies of both electrons become positive and do not take negative values during the subsequent time evolution

With this model, the formation of the fingerlike structure in correlated electron momenta was thoroughly investigated \cite{Liu2008}. Two types of classical trajectories responsible for the emergence of this structure were identified (see Ref.~\cite{Liu2008} for details). Analysis of these two types of trajectories has shown that the Coulomb focusing effect is crucially important for the emergence of electrons with high energies, and therefore for the emergence of the fingerlike structure. More specifically, this structure originates due to the electron-electron collision assisted by the nuclear attraction. It was also shown that this structure is sensitive to the relative perpendicular momentum of the two electrons, which is in agreement with experimental results \cite{Rudenko2007} and provides a credible explanation of the 1D TDSE calculations \cite{Lein2000}. 

The other major success of the trajectory-based model is the study of NSDI in the below the recollision threshold (BRT) regime, when the maximum energy of the returning electron is smaller than the ionization potential of the second bound electron. The experimental study \cite{Rudenko2008} analyses distributions of correlated electron momenta along the polarization direction at different laser intensities ranging from above to below the recollision threshold. This study revealed a transition from dominantly populated first and third quadrants of the correlated distributions to dominant population of the second and fourth quadrants with a decrease in the laser intensity for NSDI in argon. It is worth noting that two different physical mechanisms are found to be responsible for the NSDI effect: recollision-induced double ionization (RIDI) and recollision-excitation with subsequent ionization (RESI), see, e.g., Ref.~\cite{FariaRev2011}. In the RIDI mechanism, the first returning electron releases the second electron by impact ionization in the scattering process. This mechanism is implied in the three-step model \cite{Corkum1993}. In contrast to this, in the RESI mechanism, the first returning electron excites the second bound one. The second electron is then released by tunneling ionization near the next maximum of the oscillating laser field. 

The NSDI process in the BRT regime was studied in Ref.~\cite{Liu2008} by using the semiclassical model. This model extends the model described above by accounting for the RESI effect. In order to take RESI into account, the bound electron is allowed to tunnel through the potential barrier whenever it reaches the turning point of the potential barrier formed by the laser and atomic potentials. 
The corresponding tunneling rate is calculated within the WKB approximation [Eq.~(\ref{wkb})]. It was found that both the RIDI and RESI mechanism significantly contribute to the NSDI yield.
Simultaneously, the RIDI mechanism plays the main role in transition from correlation to anticorrelation patterns. For the RIDI mechanism, multiple recollisions of the rescattered electron are necessary. 
A similar conclusion was obtained by using a purely classical approach, see Ref.~\cite{Haan2008}. 

\subsection{Nonsequential double ionization of atoms by elliptically polarized fields}

\begin{figure}[h]
\centering
\includegraphics[width=0.6\textwidth]{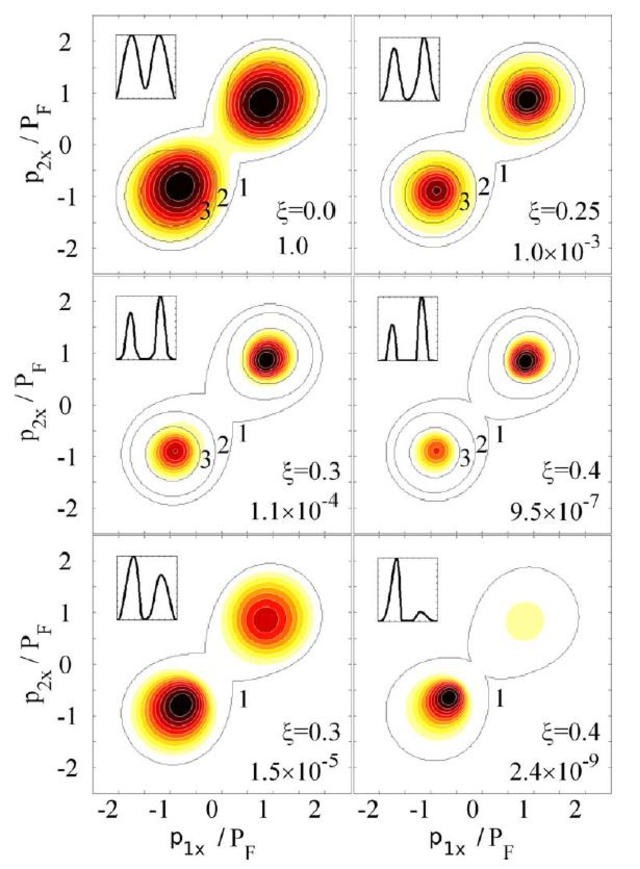}
\caption{The distribution in the $\left(p_{1x},p_{2x}\right)$ plane for $p_{1y}>0$ and $p_{2y}>0$ and various ellipticities evaluated with account of six returns of an ionized electron to its parent ion (two upper rows). Here $\vec{p}_{1}$ and $\vec{p}_{2}$ are the momenta of both electrons, and $P_F=F/\omega$. The third row presents the contribution from the first return only. The numbers in the lower right corners give the relative suppression of the larger maximum in the panel with respect to linear polarization. The boundaries of the classically allowed regions for several specified returns are exhibited and identified by the numbers next to the respective curve. The insets in the upper left corners show the cuts through the corresponding 2D distributions along the $p_{1x}=p_{2x}$ diagonal. The parameters correspond to the ionization of neon by the field of a Ti:Sa laser with intensity of $8\times10^{14}$~W/cm$^2$. From \href{https://doi.org/10.1103/PhysRevA.77.063405}{N.~I.~Shvetsov-Shilovski et al., 2008}}\label{nsdi_ellip}
\end{figure}

The first calculations of ion and electron momentum distributions in NSDI of a model atom driven by an elliptically polarized laser field were presented in Ref.~\cite{Shvetsov2008}. In that study, a simple semiclassical framework was employed, combining tunneling ionization, classical propagation of the electron in the laser field, and phase-space considerations. The electron–electron interaction was approximated by a localized three-body contact potential at the ionic core (see details in Ref.~\cite{Shvetsov2008}). It was demonstrated that when the ellipticity exceeds about 0.3, the contribution of the shortest electron trajectory—corresponding to the fastest return to the ion—ceases to dominate the double-ionization yield, in contrast to the situation under linear polarization. This change is accompanied by pronounced asymmetries in both ion momentum distributions and electron–electron correlations, as illustrated in Fig.~\ref{nsdi_ellip}. Since late electron returns are suppressed in ultrashort pulses, these asymmetries were found to depend sensitively on the pulse duration. The study also suggested that the predicted features in the electron momentum distributions should be observable experimentally, thus providing evidence for the role of late electron returns. To date, a plethora of works exploring various aspects of NSDI in elliptically and circularly polarized fields has appeared, see e.g., Refs.~\cite{Hao2009,XuWang2009,Wang2010,WangNJP2010,Xu2017,Yu2012,Wan2016,Yu2013,Chen2013,Kang2018a,Kang2018b,Kang2018c}. All these studies are based on some semiclassical models. 

Further investigation of the NSDI process induced by a field with elliptical polarization was carried out in Ref.~\cite{Hao2009} for Ne. The semiclassical model \cite{Hao2009} accounts for both the laser field and the Coulomb interaction between the nucleus, and between the two electrons. The tunnel exit point of the first electron was found from Eq.~(\ref{tunex}), and the weight of each trajectory was evaluated using a formula similar to Eq.~(\ref{tunrate}). Initial conditions of the second electron were distributed using a microcanonical distribution. The dependence of the yield of Ne$^{2+}$ ions on ellipticity calculated in \cite{Hao2009} is in good agreement with experimental data \cite{Corkum1994}.  The frequency dependence of the ratio N$^{2+}$:Ne$^+$ was investigated for different ellipticities. It is found that for high laser frequencies, this ratio increases rapidly with increasing wavelength, and it does not depend on the ellipticity. Simultaneously, for low frequencies (long wavelengths), the ratio reaches its maximum and decreases with increasing wavelength \cite{Hao2009}. In turn, the maximum of the ratio decreases with wavelength. 

In Ref.~\cite{XuWang2009}, the NSDI process in an elliptically polarized field was investigated by using the model accounting for the electron motion in both laser and soft-core Coulomb fields. The electron-electron interaction was also simulated with the soft-core Coulomb potential. While the study \cite{Shvetsov2008} was restricted to ellipticity smaller than $0.5$, momentum distributions of doubly charged ions were studied for ellipticities from $0$ to $1$ in Ref.~\cite{XuWang2009}. Therefore, the evolution of the ion momentum distributions when polarization changes from linear to circular was studied for the first time \cite{XuWang2009}.  It was shown that the contributions of sequential and nonsequential double ionization in the ion momentum distributions can be clearly distinguished.  It was found that the ellipticity of the strong laser pulse is closely related to the timing of the NSDI process \cite{Wang2010}. More specifically, the analysis of classical trajectories has shown that higher ellipticity forces the beginning of NSDI into a narrower time window. In turn, this ``pins" the field strength of the ionizing field in an unexpected way \cite{Wang2010}. By using a trajectory-based model, it was shown that the NSDI process can exist even in circularly polarized laser fields \cite{WangNJP2010}. It turns out that the underlying mechanism is again based on recollision. NSDI in a circularly polarized laser was also studied in Ref.~\cite{Xu2017} for Mg atoms. Semiclassical simulations employing the soft-core Coulomb potential have shown that for circular polarization only a single recollision event leads to the NSDI process. 

The NSDI of Ar by strong elliptically polarized laser pulses was studied in Ref.~\cite{Yu2012}. As in Ref.~\cite{XuWang2009}, the soft-core Coulomb potential was applied for the description of electron-electron interaction. It was found that, as in linearly polarized field, distributions of correlated electron momenta along the polarization direction are very sensitive to the laser intensity. Furthermore, the finger-like structure is again revealed in these distributions. The analysis of classical trajectories has shown that the final state electron-electron repulsion plays the main role in the formation of the finger-like structure \cite{Yu2012}. A closer look at the recollision process in NSDI with an elliptically polarized laser field was taken in Ref.~\cite{Wan2016}. A transition from short trajectories of recolliding electrons in linearly polarized fields to long trajectories in the fields with elliptical polarization is clearly identified. This transition can also be observed by measuring the CEP dependence of the correlated electron momentum distribution \cite{Wan2016}. 

The dependence of the NSDI process on the CEP of the elliptically polarized laser pulse was further investigated in \cite{Yu2013}. The study of Ref.~\cite{Yu2013} analyses both sequential double ionization and NSDI in the Xe atom. It is shown that electron momentum distributions generated in both the sequential and nonsequential processes strongly depend on the absolute phase of the pulse. 
The analysis of relevant classical trajectories has shown that the ionization times of both electrons in the sequential mechanism and the recollision time in the NSDI process are highly sensitive to the absolute phase \cite{Yu2013}. This explains the sensitivity of momentum distributions to the CEP.

In Ref.~\cite{Chen2013}, the distributions of correlated electron momenta from NSDI for neon in an elliptically polarized field were calculated by using a trajectory-based model. The distributions show a nontrivial behavior with increasing ellipticity that can be understood as a result of an interplay between the effects of the transverse electric field and the Coulomb potential on the motion of the first tunnel-ionized electron. It is found that for ellipticities larger than $0.2$, multiple return collision trajectories dominate in the NDSI yield and determine the correlated electron momentum distributions \cite{Chen2013}. 
 
A joint experimental and theoretical study of the NSDI process in neon induced by elliptically polarized laser pulses was performed in Refs.~\cite{Kang2018a,Kang2018b}. In the study of Ref.~\cite{Kang2018a}, the asymmetry of correlated electron momentum distributions predicted in Ref.~\cite{Shvetsov2008} was for the first time observed experimentally. Moreover, an ellipticity-dependent asymmetry of ion momentum distributions was also studied in Ref.~\cite{Kang2018b}. These asymmetry patterns were analyzed using a semiclassical model. The pattern in correlated electron distributions was attributed to the ellipticity-dependent probability distributions of the time of recollision \cite{Kang2018a}. Therefore, the correlated electron emission can be controlled, and the recolliding electron trajectories can be governed on subcycle time scales by varying the ellipticity of the laser pulse.  

Distributions of electron momentum components along the minor axis of the polarization ellipse were studied in Ref.~\cite{Kang2018c} both experimentally and theoretically. The study \cite{Kang2018c} has revealed a characteristic dependence of these distributions on the ellipticity. The semiclassical simulations have reproduced experimental observations. These simulations have shown that the dynamical energy sharing during the process of electron emission is affected by the ellipticity. Therefore, it is possible to achieve control of the energy sharing between the two emitted electrons by varying the ellipticity of the laser pulse. 

\subsection{Nonsequential double ionization of molecules}

One of the first semiclassical models aimed at the description of NSDI in molecules was developed in Ref.~\cite{Li2007} for a simple hydrogen-like molecule. This model accounts for the laser field and the Coulomb interaction between the two electrons, as well as between each of the electrons and molecular nuclei on an equal footing. Indeed, the evolution of the two-electron system is determined by the following Newton's equation of motion (\ref{newton_nsdi}) with $V^{j}_{\mathrm{ee}}=-Z_{\mathrm{eff}}/r_{aj}-Z_{\mathrm{eff}}/r_{bj}$, where $r_{aj}$ and $r_{bj}$ are the distances between the $j$-th electron and nuclei $a$ and $b$, respectively. The tunneling rate for a model hydrogen-like molecule in the quasistatic approximation was obtained in \cite{Li2007}. For the H$_2$ molecule, this formula correctly describes the dependence of the ionization probability on the orientation of the molecule. This rate, multiplied by the Gaussian exponent from Eq.~(\ref{tunrate}) describing the distribution over the initial transverse velocities is used as a weight of each classical trajectory \cite{Li2007}. 

The tunnel exit point $\mathbf{r}_{10}$ for the first electron is found from the solution of the equation:
\begin{equation}
\mathbf{r}_{10}\mathbf{F}\left(t\right)-V^{1}_{ne}-\int\frac{\left|\psi\left(\mathbf{r}\right)\right|^2}{\left|\mathbf{r}_{10}-\mathbf{r}^{~\prime}\right|}d\mathbf{r}^{~\prime}=I_{p1},
\end{equation}
where $I_{p1}$ is the ionization potential of the molecule and the wave function $\psi\left(\vec{r}\right)$ is obtained within the linear combination of atomic orbital and molecular orbital (LCAO-MO) approximation. The initial conditions of the bound electron were determined in accord with the ground-state wave function of the singly-charged molecule obtained again using the LCAO-MO approximation. The semiclassical model \cite{Li2007} was applied to the investigation of the NSDI process in diatomic molecules aligned parallel and perpendicular to the strong linearly polarized laser pulse. The alignment dependence of the NSDI process in the N$_2$ molecule calculated within the model is in qualitative agreement with the experimental results \cite{Li2007}. 

The model \cite{Li2007} was further developed in Refs.~\cite{LiuLiu2007,Ye2008}. First, the initial positions and momenta of the bound electron are distributed by using the single-electron microcanonical distribution \cite{Abrines1966}. Second, the model \cite{LiuLiu2007,Ye2008} is able to describe NSDI in the barrier suppression regime. This is achieved by distributing the initial conditions with double-electron microcanonical distribution (see Ref.~\cite{Meng1989}). The model developed in Refs.~\cite{LiuLiu2007,Ye2008} is able to reproduce experimental results for laser intensities ranging from the tunneling to the barrier suppression regime. It has predicted a substantial alignment effect in the ratio of double- over single-ion yield. The analysis of classical trajectories can explain and visualize subcycle dynamics of the molecular double ionization \cite{Ye2008}.

Semiclassical modelling has also been used to study molecular NSDI in Refs.~\cite{Emma2009,Emma2011}. The study of Ref. [4] focuses on the NSDI process induced by a short ($6$ fs) laser pulse. Laser intensities corresponding to both tunneling and barrier suppression regimes (close to the threshold) were considered \cite{Emma2009}. In the tunneling regime, two new mechanisms of double ionization have been revealed in Ref.~\cite{Emma2009}. The semiclassical simulations have revealed an anticorrelation pattern in distributions of electron momentum components along the polarization direction. This pattern is similar to the one experimentally observed for NSDI in Ar below the recollision threshold \cite{Rudenko2008}. In contrast to this, relatively long laser pulses ($27$ fs) were employed in \cite{Emma2011}. Again, laser intensities ranging from the tunneling to the barrier suppression regime were used in simulations. It is found that for intermediate intensities in the barrier-suppression regime, a characteristic square pattern prevails in the distributions of correlated electron momenta. The square pattern emerges due to the contribution of the delayed pathway of NSDI. This pathway corresponds to the situation where one of the electrons is ejected with a delay after the recollision. It is shown that for intermediate intensities the delayed pathway is dominated by the ``soft" recollision.

\subsection{Recent developments in semiclassical simulations of nonseqeuntial double ionization}

In recent years, semiclassical models have been used for studies of more complex aspects and features of NSDI. Here, we only briefly mention a few recent examples of applications of semiclassical models to the study of NSDI.  One of them is the nondipole effect in the NSDI process, see, e.g., Refs.~\cite{Emma2017e,Emma2017f}. The trajectory-based model fully accounting for nondipole effects in the electron dynamics was applied to double ionization of He and Xe atoms driven by a near-infrared and mid-infrared laser pulses, respectively. It was shown that if the nondipole effects are accounted for, the sum of the electron momenta is an order of magnitude larger than twice the average electron momentum in the polarization direction of the laser field in single ionization. This unexpectedly large value of the sum of the momenta is found to be due to strong recollisions \cite{Emma2017e}. For linearly and slightly elliptically polarized laser fields, the combination of the magnetic field and recollisions acts as a gate \cite{Emma2017f}. The gate favors initial momenta of the tunneling electron that are more transverse to the electric field and are opposite to the propagation direction of the laser field.  It is shown that the asymmetry in the transverse initial momentum leads to an asymmetry in observables of NSDI. 

The trajectory-based model has been successfully applied to the study of NSDI induced by a near-single cycle laser pulses \cite{Chen2017}. Asymmetry parameters, distributions of the sum of the two electron momentum components along the polarization direction, and distributions of correlated electron momenta calculated for various intensities and CEPs are shown to be in a good agreement with results of kinematically complete experiments. Semiclassical simulations have made it possible to identify contributions of direct and delayed pathways (when one electron ionizes with a time delay after recollision) of double ionization in observables (see Ref.~\cite{Chen2017} for details). Finally, it is shown that an experimentally observed anti-correlation pattern in distributions of momentum components along the laser polarization direction emerges due to a transition from strong to soft recollisions with increasing intensity. 

Recently, semiclassical simulations have shown that RESI is not always the main mechanism behind the delayed pathway of NSDI \cite{Emma2008}. It was found that for He atoms irradiated by near-single-cycle pulses with the wavelength of 400~nm and the intensity below the recollision threshold, another mechanism overtakes RESI in the delayed NSDI pathway \cite{Emma2018f}. In this mechanism, the first returning electron transfers part of its energy to the second bound electron and excites it. The second electron undergoes a Coulomb slingshot motion due to the attractive force of the nucleus, and it is released from the atom mostly near the second extremum of the laser field with the assistance of both the laser and the nucleus. Therefore, this mechanism was called slingshot-NSDI. Within the slingshot-NSDI, the two liberated electrons escape in opposite directions along the laser polarization. It should be stressed that the anticorrelated pattern had been previously attributed only to multiple recollisions within the RESI mechanism. The slingshot-NSDI was found to dominate at low laser intensities. It was shown that the slingshot-NSDI has very distinct fingerprints in momentum distributions of the two electrons, see Ref.~\cite{Emma2019}. The corresponding features can be observed in an experiment. 

Semiclassical simulations have been used to explore the application of the streaking technique (see Ref.~\cite{Landsman2024}) to strong-field double ionization \cite{Emma2019a}. It was shown that both sequential and nonsequential double ionization processes result in streaking delays that differ from each other and from those for single ionization. The simulation results agree well with experimental data. 

The ECBB model has recently been used to investigate the triple ionization of Ne atoms \cite{Emma2023}. The calculated distributions of the sum of the final electron momenta generated in triple ionization are in good agreement with experiment. By using the semiclassical model, the main features of these momentum distributions were explained in terms of the main pathways of three-electron escape in Ne. It is shown that these pathways are different for Ne and Ar, and, as a result, electron momentum distributions are different for these two atoms. 

Nondipole effects in triple ionization of Ne by infrared pulses at intensities where electron-electron correlations play a major role were investigated in Ref.~\cite{Emma2023a}. Application of the ECBB model has made it possible to identify a prominent signature of nondipole effects. It is found that the component of the average sum of the final electron momenta in the direction of pulse propagation is large and positive (see Ref.~\cite{Emma2023a} for details). This positive momentum offset is absent in the dipole approximation. It is found that the offset arises mainly from the change in momentum due to the magnetic field.

\section{Conclusions and outlook}
\label{conc}

Semiclassical models that use Newton’s equations to describe the motion of an electron after it has been promoted to the continuum are a powerful tool in strong-field, ultrafast, and attosecond physics. Very often these models provide a better understanding of the physical mechanism underlying the strong field phenomenon under study than other theoretical approaches used in this area of physics, such as the direct numerical solution of the TDSE and SFA. Indeed, the semiclassical models operate with classical trajectories that are easy to imagine and visualize on a computer screen or a sheet of paper. The calculation of a classical trajectory involves obtaining not only the coordinates, but also the velocity and acceleration of the electron at any time. In combination with quantum data, such as the ionization rate, initial velocity, and the exit from under the potential barrier, this information is often sufficient to explain the subtle features of strong field phenomena. Another important advantage of semiclassical models is that in many cases they are computationally simple especially compared to the solution of the TDSE. Even if this is not the case, trajectory-based calculations can be easily parallelized, making them particularly suitable for computer clusters. 

Here we discuss the development and application of semiclassical models to the cornerstone processes of strong-field physics: ATI, HHG, NSDI, and FTI. We show how ensembles of classical trajectories being properly initialized, calculated, and corrected for quantum effects make it possible to obtain observables characterizing these strong-field processes. We also see how exactly the semiclassical modeling provides a unique opportunity to uncover complex physical nature of various phenomena induced by strong laser fields. For the NSDI process alone, this concerns the physical mechanisms underlying this process at different laser intensities, the origin of the characteristic fingerlike structure in electron-electron correlations, the role of long and short trajectories in an elliptically polarized laser field, and the role of the nondipole effects. Other well-known examples include Coulomb effect in the interference patterns of electron momentum distributions generated in the ATI process, different pathways of FTI and role of nondipole effects in this process - these are only some of the problems addressed in the present review.   

Naturally, the applications of trajectory-based models to highly nonlinear quantum phenomena of strong-field physics involve a number of difficulties. It is clear that these difficulties may be different for various laser-induced phenomena. For ATI, accounting for quantum interference effects remained a problem until the advent of the TCSFA, QTMC, CQSFA, and SCTS in the last decade. Nevertheless, these models aimed at the description of interference patterns in electron momentum distributions are often computationally demanding since they either operate with large ensembles of classical trajectories (TCSFA, QTMC, and SCTS) or require solving the inverse problem (CQSFA and a recent implementation of the SCTS model). This problem is particularly pressing in semiclassical simulations of the high-order ATI. The main problem with the semiclassical description of HHG is that there is no recombination process if a returning electron is described classically. The hybrid quantum-classical approaches used to bypass this fact may have their own limitations. For both high-order ATI and HHG, a thorough comparison of predictions of trajectory-based models with the TDSE results is required. It is clear that correlated electron dynamics is crucial for the NSDI process. Simultaneously, it is hard to capture such dynamics in a semiclassical model. Although classical Coulomb interactions can capture electron-electron correlation to some extent, quantum correlations including exchange effects and entanglement are entirely missing in a trajectory-based description. The challenges in description of the FTI process come from the need for the application of a classical criterion for the capture into Rydberg states. This classical criterion may not accurately describe the quantum nature of the electron trapping involving discrete energy levels and coherence effects. 

Simultaneously, the proper choice of initial conditions, i.e., starting points of classical trajectories and the corresponding initial velocities, is important for an accurate modelling of all these phenomena. The outcomes of trajectory-based simulations are often quite sensitive to the initial conditions. It is not an exaggeration to say that initial conditions are a weak point of the semiclassical approach to strong-field phenomena. Indeed, these initial data for trajectory-based simulations are the interface between a quantum and classical approach, and they can be considered as a quantum input of a semiclassical model.

It is therefore natural to expect that further development of trajectory-based models will be aimed at an improved description of the tunneling step. The application of nonadiabatic tunneling models and quasiprobability distributions are promising steps in this direction. There is also a growing interest in using machine learning and data-driven approaches to classify trajectories, optimize initial conditions, and calculate large ensembles of classical trajectories efficiently.
On the other hand, semiclassical modelling of multielectron dynamics, which is not discussed here, has been also attracting considerable attention. These perspective directions are shaping the development of semiclassical models for strong-field physics There is every reason to believe that semiclassical modelling will remain an invaluable tool in strong-field physics finding a balance between physical insight and computational efficiency. 

\section*{References}

\bibliography{paper-1} 

\end{document}